\documentclass[preprint,preprintnumbers,amsmath,amssymb]{revtex4}

\usepackage{graphicx}\usepackage{bm}
\usepackage{color}
\begin{document}
\title{
{\it \textbf{Ab initio}} prediction of an order-disorder transition in Mg$_2$GeO$_4$: implication for the nature of super-Earth's mantles
}

\author{Koichiro Umemoto$^{1,2}$, and Renata M. Wentzcovitch$^{3,4}$}

\affiliation{
${}^1$Earth-Life Science Institute, Tokyo Institute of Technology, Tokyo, Japan \\
${}^2$Theoretical Quantum Physics Laboratory, Cluster for Pioneering Research, RIKEN, Wako-shi, Saitama 351-0198, Japan\\
${}^3$Department of Applied Physics and Applied Mathematics, Columbia University, New York, NY, USA.\\
${}^4$Department of Earth and Environmental Sciences, Lamont Doherty Earth Observatory, 
Columbia University, New York, NY, USA.
}
\date{\today}
\begin{abstract}
Here we present an {\it ab initio} prediction of an order-disorder transition (ODT) from $I\bar{4}2d$-type to a Th$_3$P$_4$-type phase in the cation sublattices of Mg$_2$GeO$_4$, a post-post-perovskite (post-PPv) phase. This uncommon type of prediction is achieved by carrying out a high-throughput sampling of atomic configurations in a 56-atom supercell followed by a Boltzmann ensemble statistics calculation. Mg$_2$GeO$_4$ is a low-pressure analog of $I\bar{4}2d$-type Mg$_2$SiO$_4$, a predicted major planet-forming phase of super-Earths' mantles. Therefore, a similar ODT is anticipated in $I\bar{4}2d$-type Mg$_2$SiO$_4$ as well, which should impact the internal structure and dynamics of these planets. The prediction of this Th$_3$P$_4$-type phase in Mg$_2$GeO$_4$ enhances further the relationship between the crystal structures of Earth/planet-forming silicates and oxides at extreme pressures and those of rare-earth sesquisulfides at low pressures.
\end{abstract}

\maketitle

\section*{Introduction}

{\it Ab initio} quasiharmonic (QHA) calculations of polymorphic phase transitions at extreme pressure and temperature conditions have proven to be highly predictive for nearly two decades \cite{RIMG2010}. Combined with materials discovery methods (e.g., Refs.~\cite{Glass2006,Lonie2011,Pickard2011,Wang2012,Wu2014,Curtis2018}), these simulation tools offer a powerful approach to investigating phase transition phenomena at challenging experimental conditions typical of planetary interiors. Mineral physics and geophysics have benefitted immensely from these developments in materials simulations in the past two decades. For example, in 2004, {\it ab initio} predictions played a crucial role in discovering and elucidating the major post-perovskite (PPv) transition in MgSiO$_3$ bridgmanite at deep Earth interior conditions, e.g., 2,500K at 125 GPa \cite{Murakami2004,Oganov2004,Tsuchiya2004}.
More recently, {\it ab initio} QHA calculations have explored pressure and temperature conditions expected in interior of super-Earths, terrestrial-type exoplanets more massive than Earth. The interest in these planets stems from their similarities and differences with Earth and their potential habitability. Besides, {\it ab initio} methods are highly predictive when addressing planet-forming silicates and oxides, motivating experiments. These phases are oxides involving Mg, Si, ferrous and ferric Fe, Al, and Ca, primarily. In deep interiors of large super-Earths with up to $\sim$13 Earth mass ($M_\oplus$), the range of pressures and temperatures can reach tens of tera-Pascals (TPa) (hundreds of Mbar) and $10^4$--$10^5$ K \cite{Hakim2018,vandenBerg2019}. Despite remarkable developments in experimental techniques \cite{Smith2014,Dubrovinsky2012,Dubrovinskaia2016,Dewaele2018,Sakai2018}, these pressure-temperature conditions are still very challenging to experiments making {\it ab initio} predictions of phase transition phenomena in planet-forming phases critical to advancing planetary modeling. 
This progress has been registered in a series of {\it ab initio} discoveries concerning the nature of super-Earth's mantle-forming phases in the important MgO-SiO$_2$ system \cite{Wu2014,Umemoto2006a,UmemotoWentzcovitch2011,Niu2015,Umemoto2017}. They have culminated in partial experimental confirmations in a low-pressure analog system, NaF-MgF$_2$ \cite{Dutta2018}, and detailed modeling of these planets' internal structure and dynamics \cite{Hakim2018,vandenBerg2019}.  In particular, a sequence of ``post-post-perovksite'' (post-PPv) transitions in the MgO-SiO$_2$ system \cite{Wu2014,Umemoto2006a,UmemotoWentzcovitch2011,Niu2015,Umemoto2017} was predicted to occur up to $\sim$3 TPa and 10,000 K, starting from MgSiO$_3$ PPv and ending in its dissociation into the elementary oxides MgO and SiO$_2$. This dissociation process was predicted to occur in three stages: 1) a dissociation reaction, MgSiO$_3$ PPv $\to$ $I\bar{4}2d$-type Mg$_2$SiO$_4$ + $P2_1/c$-type MgSi$_2$O$_5$; 2) further dissociation of $P2_1/c$-type MgSi$_2$O$_5$ $\to$ $I\bar{4}2d$-type Mg$_2$SiO$_4$ + Fe$_2$P-type SiO$_2$; 3) final dissociation of $I\bar{4}2d$-type Mg$_2$SiO$_4$ $\to$ CsCl-type MgO + Fe$_2$P-type SiO$_2$. If MgSiO$_3$ coexists with MgO or SiO$_2$, other intermediate recombination reactions producing Mg$_2$SiO$_4$ or MgSi$_2$O$_5$ can intervene in the three-stage dissociation process. These post-PPv transitions were shown to have profound effects on super-Earths' mantle dynamics \cite{Hakim2018,vandenBerg2019}. However, the exceedingly high predicted transition pressures ($>\sim500$ GPa) make their experimental confirmation quite challenging. 

Two potential low-pressure analog systems, i.e., MgO-GeO$_2$ and NaF-MgF$_2$, were proposed as viable experimental alternates \cite{UmemotoWentzcovitch2015,UmemotoWentzcovitch2019}. Both displayed some of these novel high-pressure phases found in the MgO-SiO$_2$ system. NaMgF$_3$ PPv was predicted to exhibit the novel $P2_1/c$-type structure of MgSi$_2$O$_5$ under pressure before its full dissociation into NaF and MgF$_2$. The predicted phases and transformations were experimentally confirmed \cite{Dutta2018}. In contrast, MgGeO$_3$ PPv under pressure was expected to produce the novel $I\bar{4}2d$-type Mg$_2$SiO$_4$ structure.
The predicted reactions in the MgO-GeO$_2$ system have not been experimentally confirmed yet since they happen at higher pressures. Also, both systems were expected to display the respective analog recombination reactions. This state of affairs brings us to our present study. 

\begin{figure}[h]
\hbox to \hsize{\hfill
\includegraphics[width=100mm]{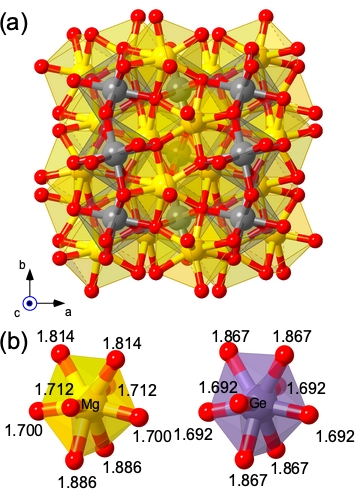}
\hfill
 }
\caption{
(a) Crystal structure of $I\bar{4}2d$-type Mg$_2$GeO$_4$. Yellow, grey, and red spheres denote Mg, Ge, and O ions, respectively. 
(b) Coodination polyhedra around Mg and Ge ions. Numbers next to the O atoms represent Mg-O and Ge-O bond-lengths in Angstroms at 400 GPa.
}
\label{structure}
\end{figure}

Using {\it ab initio} techniques, we predict another type of phase transition in the MgO-GeO$_2$ system, a temperature-induced change from $I\bar{4}2d$-type to Th$_3$P$_4$-type structure in Mg$_2$GeO$_4$. This is not a regular polymorphic phase transition but an order-disorder transition (ODT) in the cation sublattices. 
The crystal structure of $I\bar{4}2d$-type Mg$_2$GeO$_4$ is shown in Fig.~\ref{structure}. In this phase, Mg and Ge cations are regularly ordered. Disorder in the cation sublattices makes the cation sites indistinguishable and results in the Th$_3$P$_4$-type structure. Details of both crystal structures are given in Supplementary Information.
The result predicted here is unexpected since disorder occurs between two sublattices containing cations with nominally different valences. Specifically, Mg and Ge are known to form stoichiometric end-member phases (MgO and GeO$_2$), preserving their nominal valence at comparable pressures. This type of prediction is also uncommon since it requires reliable methods to compute free energy in disordered solid-solutions, which is computationally much more costly than regular polymorphic transition. The predicted pressure and temperature transition conditions for this ODT are more easily achievable in laboratory experiments than the analog one in Mg$_2$SiO$_4$. Still, a similar ODT is also expected to occur in $I\bar{4}2d$-type Mg$_2$SiO$_4$ and should be crucial for modeling interiors of super-Earths.

\section*{Statistical treatment}

The ODT critical temperature, $T_c$, was calculated using the same approach previously used to compute the ice-VII to -VIII ODT boundary \cite{Umemoto2010}. The method consists in sampling, if not all, the most significant symmetrically distinct (irreducible) atomic configurations of a chosen supercell. To represent the disordered Th$_3$P$_4$-type phase we chose a 56-atom supercell (8 Mg$_2$GeO$_4$, $\sqrt{2} \times \sqrt{2} \times 1$ supercell of the conventional unit cell) and generated an ensemble of 125 irreducible configurations using the scheme described below. We then computed the static partition function for this ensemble of 125 configurations:  
\begin{equation}
Z_{static}(V,T)=\sum^{125}_{i=1} w_i \exp \left( -\frac{E_i(V)}{k_BT} \right),
\label{partition}
\end{equation}
where $E_i(V)$ and $w_i$ are the total energy and multiplicity of the $i$th irreducible configuration ($\sum_{i=1}^{125}w_i=3489$), and $k_B$ is the Boltzmann's constant. 
The static partition function is then extended to include zero-point motion (ZPM) and phonon thermal excitation energies within the QHA \cite{Umemoto2010,Qin2019}. From $Z(V,T)$, all thermodynamic potentials and functions can be calculated: Helmholtz free energy $F(V,T)=-k_B\ln Z$, pressure $P(V,T)=-(\partial F/\partial V)_T$, Gibbs free energy $G(V,T)=F+PV$ (converted to $G(P(V,T),T)$), entropy $S(V,T)=-(\partial F/\partial T)_V$, constant-volume heat capacity $C_V(V,T)=-T(\partial ^2 S/\partial T^2)_V$, constant-pressure heat capacity $C_P(P,T)=-T(\partial ^2 S/\partial T^2)_P$, and so forth. Finally the ODT is obtained by locating a peak in $C_P{(T)}$.
The procedure used to generate the 125 cation configurations in Eq.~\ref{partition} is schematically depicted in Fig. S1. We started with the ordered structure containing the 24 cations (16 Mg and 8 Si ions) in their respective Wyckoff sites of the $I\bar{4}2d$-type phase. This lowest enthalpy configuration is shown in Fig.~\ref{structure}(a) and corresponds to the leftmost configuration in Fig. S1. We refer to it as the ``zero-interchange structure''. Then we sequentially interchanged Mg/Ge pairs once. These single interchanges produce 128 (16 $\times$ 8) configurations where only 4 are irreducible, each with its distinct multiplicity. We refer to this first generation of structure as ``one-interchange'' configurations. Starting from these 128 configurations, we repeat this process and produce 3360 configurations among which 120 are irreducible ``two-interchange'' configurations. With zero, one, and two cation interchanges, a total of $1 + 4 + 120 = 125$ configurations were generated and used to compute the partition function.
Discussion of convergence issues related to supercell size and number of configurations is offered in the Supplementary Information section.

\section*{Results and discussion}

\begin{figure}[h]
\hbox to \hsize{\hfill
\includegraphics[width=120mm]{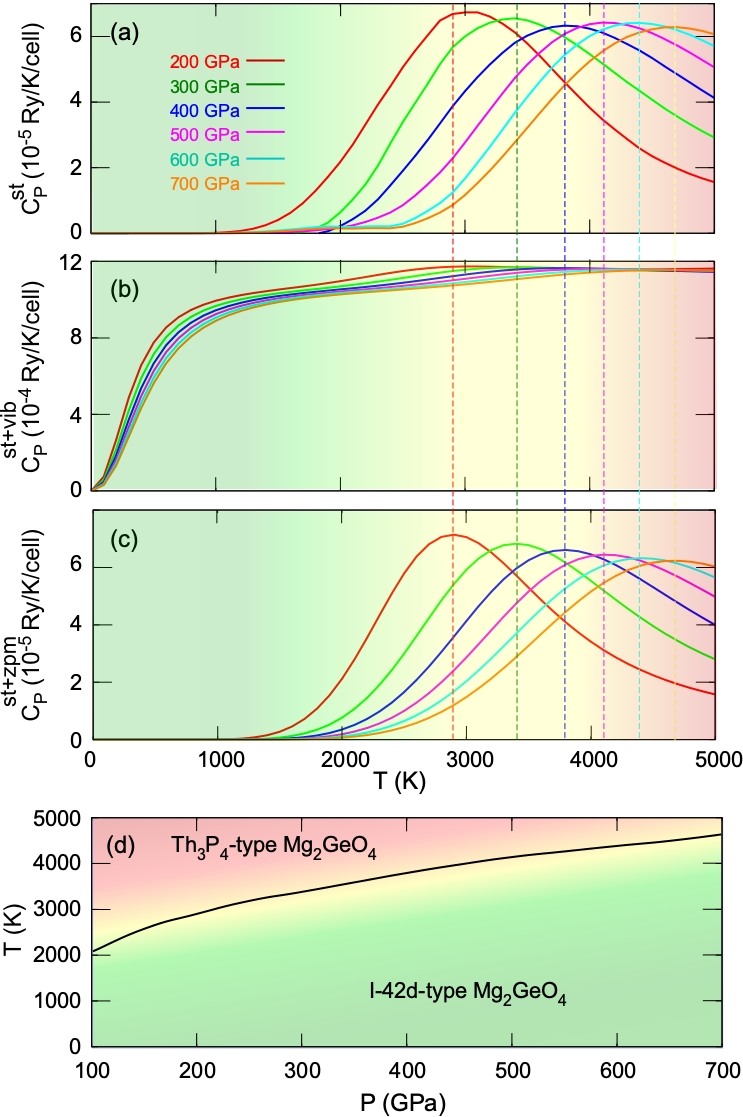}
\hfill
 }
\caption{
Constant-pressure heat capacities, $C_P$, calculated using 125 configurations in the 56-atom supercell and (a) static free energies only (no vibrational effects included) ($C_P^{st}$), (b) free energies calculated using the QHA ($C_P^{st+vib}$), and (c) static free energy plus $E_{ZPM}$ ($C_P^{st+zpm}$). Vertical dashed lines indicate the peak temperatures of $C_P^{st+zpm}$. (d) Phase boundary of the ODT between $I\bar{4}2d$-type and Th$_3$P$_4$-type Mg$_2$GeO$_4$ obtained from the peak temperatures in (c).
}
\label{Cp}
\end{figure}

$C_P$ profiles at several pressures obtained using static free energy calculations ($C_P^{st}$) are shown in Fig.~\ref{Cp}(a). The peak in $C_P^{st}$ can be easily identified. Including the vibrational free energy contribution to the total free energy, a Debye-like contribution is added to $C_P^{st}$ producing $C_P^{st+vib}$. The peak in $C_P^{st+vib}$ appears as a hump in the Dulong-Petit regime of $C_P^{st+vib}$ (Fig.~\ref{Cp}(b)). The peak would have appeared very sharp if the calculation could have been carried out with an infinitely large supercell and number of configurations. By adding only the ZPM energy, $E_{ZPM}$, the heat capacity ($C_P^{st+zpm}$) still resembles $C_P^{st}$, except that $T_c$ lowers by $\sim$100 K at 200 GPa (Fig.~\ref{Cp}(c)). This temperature shift is essentially a volume effect caused by the expansion of the equilibrium volume upon inclusion of $E_{ZPM}$.

With increasing pressure, $T_c$, i.e., the peak temperature in $C_P$, increases producing the phase boundary shown in Fig.~\ref{Cp}(d). This happens because the enthalpy difference between all configurations and the ground state one, the ordered $I\bar{4}2d$-type phase, increases with pressure as shown in Fig.~S3. Among the 125 configurations, the $I\bar{4}2d$-type phase has the lowest enthalpy at all pressures investigated here. The four irreducible configurations generated by one-interchange of cations are more similar to the ordered ground state structure and tend to have lower enthalpies than the 120 irreducible configurations produced by two-interchanges, with some exceptions. 
This behavior of $T_c$, i.e., the positive Clapeyron slope is opposite to that observed in the ice-VII to -VIII ODT. In the case of ice, all configuration enthalpies converge to a single value under pressure, that of ice X. This is because the ice ODT precedes and turns into a hydrogen-bond symmetrization transition under pressure \cite{Umemoto2010}.  Besides, in the present ODT $T_c$ is only slightly altered by quantum effects, e.g, ZPM or thermal excitation effects. It takes place above the Debye temperature, $\theta_{\rm Debye}$, in the Dulong-Petit regime of $C_V$. Therefore, it has a classical origin and can be reasonably well addressed using the static partition function in Eq.~\ref{partition}.

\begin{figure}[h]
\hbox to \hsize{\hfill
\includegraphics[width=120mm]{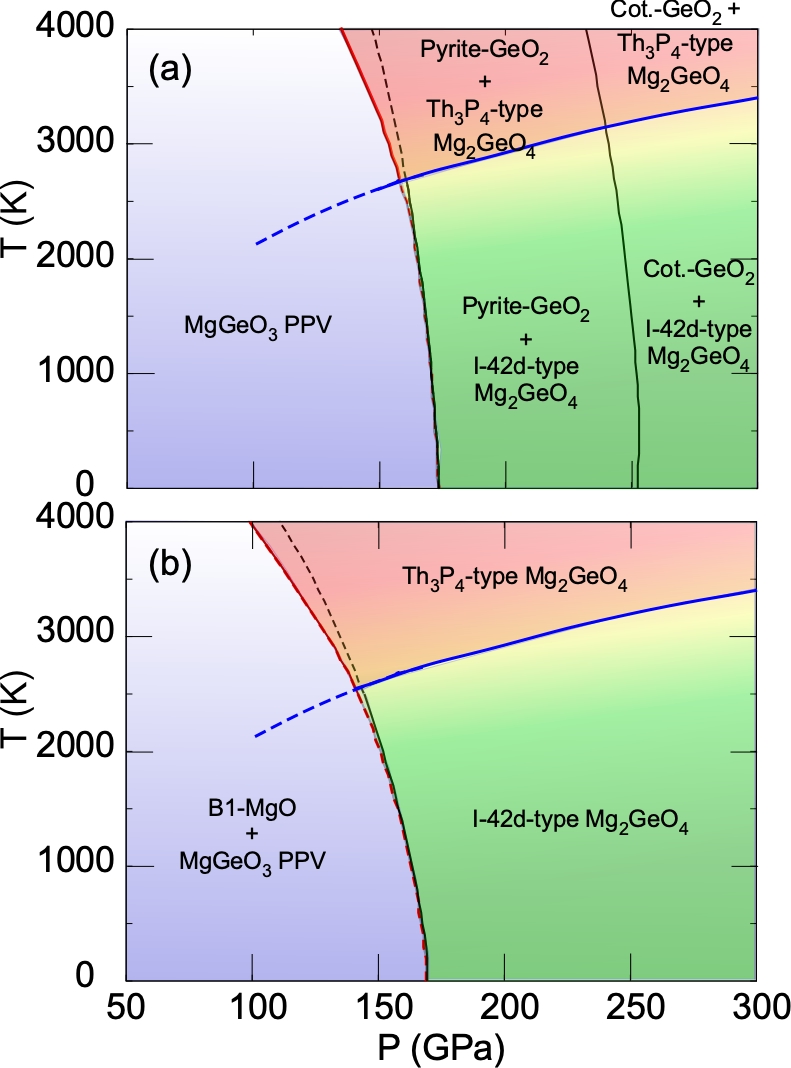}
\hfill
 }
\caption{
Phase transition boundaries in the MgO-GeO$_2$ system.
The blue solid lines show the ODT in Mg$_2$GeO$_4$ given by the peak temperature in $C_P$ (Fig.~\ref{Cp}(c)).
The red lines denote the post-PPv phase boundaries involving Th$_3$P$_4$-type Mg$_2$GeO$_4$, while the black lines denote those involving $I\bar{4}2d$-type Mg$_2$GeO$_4$ calculated in Ref.~\cite{UmemotoWentzcovitch2019}. The dashed lines represent the metastable continuation of these phase boundaries.
}
\label{PB}
\end{figure}

The following dissociation and recombination post-PPv transitions were predicted in the MgO-GeO$_2$ system ~\cite{UmemotoWentzcovitch2019}: 
\begin{description}
\item[Dissociation -] MgGeO$_3$ (PPv) $\to$ Mg$_2$GeO$_4$ ($I\bar{4}2d$-type) + GeO$_2$ (pyrite-type) at $\sim$ 175 GPa followed by the transition of GeO$_2$ from pyrite- to cotunnite-type, which is not affected by the order-disorder transition;
\item[Recombination -] MgGeO$_3$ (PPv) + MgO (B1-type) $\to$ Mg$_2$GeO$_4$ ($I\bar{4}2d$-type) at $\sim$ 173 GPa.
\end{description}
These transitions remain valid at low temperatures, but above the ODT's, $T_c$ one should replace the $I\bar{4}2d$-type phase by the Th$_3$P$_4$-type one.
The newly computed phase boundaries for these transitions are shown in Fig.~\ref{PB}. Both reactions have negative Clapeyron slopes, a common behavior in pressure induced structural transitions involving an increase in cation coordination \cite{Navrotsky1980}. The ODT widens the stability fields of the dissociation products of MgGeO$_3$ PPv (GeO$_2$ and Mg$_2$GeO$_4$) and of the recombination product (Mg$_2$GeO$_4$) because the configuration entropy lowers the Gibbs free energy of Mg$_2$GeO$_4$ in the disordered Th$_3$P$_4$-type phase at higher temperatures. The configuration entropy also decreases in magnitude the negative Clapeyron slopes of dissociation and recombination transitions $(dT/dP=\Delta V/\Delta S)$ at high temperatures. As pointed out earlier \cite{UmemotoWentzcovitch2019}, hysteresis might prevent experimental observation of these dissociation and recombination transitions from MgGeO$_3$ PPv, in which case a polymorphic from PPv to a Gd$_2$S$_3$-type phase might intervene. A similar phenomenon was observed experimentally in the NaF-MgF$_2$ system in which NaMgF$_3$ PPv transformed to a U$_2$S$_3$-type post-PPv phase at low temperatures \cite{Dutta2018}.

\section*{Conclusion}
We have predicted an order-disorder transition (ODT) in the cation sublattices of the $I\bar{4}2d$-type Mg$_2$GeO$_4$, a post-PPv phase in the Mg-Ge-O system. This type of prediction is uncommon and is not accomplished simply using modern materials discovery techniques (e.g., Refs.~\cite{Glass2006,Lonie2011,Pickard2011,Wang2012,Wu2014,Curtis2018}). In addition to structural prediction, the ODT prediction requires effective statistical sampling of atomic configurations \cite{Umemoto2010}. Besides, $T_c$ cannot be calculated by direct comparison of Gibbs free energy as in a regular first order transition. Instead, $T_c$ is obtained by calculating the position of a peak in $C_P(T)$ throughout this second order transition. In this study, this was accomplished using {\it ab initio} quasiharmonic (QHA) calculations on a 56-atom supercell. Although anharmonic effects may play a role in first order transitions at the high temperatures investigated here, harmonic or anharmonic vibrational effects play a secondary role in the present ODT.

The predicted $I\bar{4}2d$-type to Th$_3$P$_4$-type phase change expands toward lower pressures and temperatures the stability fields of the post-PPv dissociation/recombination products containing Mg$_2$GeO$_4$. The MgO-GeO$_2$ system is a partly low-pressure analog of the Earth/planet-forming MgO-SiO$_2$ system. Both $I\bar{4}2d$-type Mg$_2$GeO$_4$ and Mg$_2$SiO$_4$ were predicted to occur as post-PPv dissociation/recombination transition products. Therefore, the present study strongly suggests that a similar ODT should also occur in the Mg and Si cation sublattices of $I\bar{4}2d$-type Mg$_2$SiO$_4$ at the high temperatures typical of the deep interiors of super-Earths \cite{Hakim2018,vandenBerg2019}.

Finally, it should be noted that the Th$_3$P$_4$-type structure is a high temperature form of several rare-earth sesquisulfides, $RR'$S$_3$ ($R$,$R$'=lanthanoid or actinoid), with vacancies at cation sites \cite{Zachariasen1949,Flahaut1979}. For example, Gd$_2$S$_3$ is stabilized in the orthorhombic $\alpha$ phase at low temperature and transforms to the Th$_3$P$_4$-type $\gamma$ phase with vacancies at high temperature. As pointed out earlier \cite{Umemoto2008}, the $RR'$S$_3$ family of structures form an analog system to high-pressure phases of MgSiO$_3$ and Al$_2$O$_3$ (bridgmanite, PPv, and U$_2$S$_3$-type). Hence, the prediction of the Th$_3$P$_4$-type phase in this study strengthens further the structural relationship between Earth/planet-forming phases at ultrahigh pressures and rare-earth sesquisulfides. 

\section*{Computational details}

Calculations for the 125 cation configurations were performed using the local-density approximation \cite{Perdew1981} to density-functional theory. For all atomic species, Vanderbilt-type pseudopotentials \cite{Vanderbilt1990} were generated.
The valence electron configurations and cutoff radii for the pseudopotentials were $2s^22p^63s^2$ and 1.6 a.u. for Mg, $4s^24p^13d^{10}$ and 1.6 a.u. for Ge, and $2s^22p^4$ and 1.4 a.u. for O, respectively.
Cutoff energies for the plane-wave expansion are 70 Ry. The $2\times2\times2$ {\bf k}-point mesh was used for the 56-atom supercell.
For structural optimization under arbitrary pressures between 100 and 800 GPa, we used the variable-cell-shape damped molecular dynamics \cite{Wentzcovitch1991,Wentzcovitch1993}. Dynamical matrices were calculated on the $2\times2\times2$ {\bf q} mesh using density-functional perturbation theory \cite{Giannozzi1991,Baroni2001}. The vibrational contribution to the partition function was taken into account within the quasi-harmonic approximation (QHA) \cite{Wallace1972,Umemoto2010,Qin2019}. The $8\times8\times8$ {\bf q}-point mesh was used for the QHA summation.
All calculations were performed using the Quantum-ESPRESSO \cite{Giannozzi2009} and the \texttt{qha} software \cite{Qin2019}.

\section*{Acknowledgments}

K.U. acknowledges support of a JSPS Kakenhi Grant \# 17K05627. R.M.W. acknowledges support of a US Department of Energy Grant DE-SC0019759. All calculations were performed at Global Scientific Information and Computing Center and in the ELSI supercomputing system at the Tokyo Institute of Technology, the HOKUSAI system of RIKEN, and the Supercomputer Center at the Institute for Solid State Physics, the University of Tokyo.

\newpage
{\Large \bf Figures}

\end{document}


\title{SUPPLEMENTAL INFORMATION:\\
{\it \textbf{Ab initio}} prediction of an order-disorder transition in Mg$_2$GeO$_4$: implication for the nature of super-Earth's mantles
}

\author{Koichiro Umemoto$^{1,2}$, and Renata M. Wentzcovitch$^{3,4}$}

\affiliation{
${}^1$Earth-Life Science Institute, Tokyo Institute of Technology, Tokyo, Japan \\
${}^2$Theoretical Quantum Physics Laboratory, Cluster for Pioneering Research, RIKEN, Wako-shi, Saitama 351-0198, Japan\\
${}^3$Department of Applied Physics and Applied Mathematics, Columbia University, New York, NY, USA.\\
${}^4$Department of Earth and Environmental Sciences, Lamont Doherty Earth Observatory, 
Columbia University, New York, NY, USA.
}
\date{\today}

\maketitle

\section*{Crystal structures of Mg$_2$GeO$_4$}
The crystal structure of $I\bar{4}2d$-type Mg$_2$GeO$_4$ is shown in Fig.~1. As far as we know, this structure has not been identified experimentally so far. This phase is body-centered-tetragonal and its space group is $I\bar{4}2d$. In the $I\bar{4}2d$-type phase, both Mg and Ge cations are regularly ordered; Mg and Ge ions are at $8d$ and $4b$ Wyckoff positions, respectively. Local atomic environments around the Mg and Ge sites are very similar, both being surrounded by eight O ions (Fig.~1(b)). The eight-fold coordinated polyhedra are somewhat intermediate between NaCl-type octahedra and CsCl-type cubes. The $I\bar{4}2d$-type structure is related to that of Zn$_2$SiO$_4$-II [S1].
While the cation arrangements in Zn$_2$SiO$_4$-II are identical to those of $I\bar{4}2d$-type Mg$_2$GeO$_4$, the anion arrangement differs. In Zn$_2$SiO$_4$-II, cations are surrounded by four O ions. Average Mg-O and Ge-O bondlengths are 1.788 \AA~ and 1.780 \AA~ at 400 GPa, respectively. This similarity suggests that only a relatively small elastic strain energy would be generated by disordering Mg and Ge ions and configuration entropic effects could stabilize a disordered phase in Mg$_2$GeO$_4$ at high temperatures. Such disordered phase is cubic and its space group is $I\bar{4}3d$, a supergroup of $I\bar{4}2d$. The structure of this disordered phase is identical to that of Th$_3$P$_4$. In this structure, cations are in the $12a$ (3/8, 0, 1/4) Wyckoff sites. 

\section*{Convergence issues}
Here we test the convergence of $T_c$ with respect to the supercell size and the number of configurations. For this purpose only, we calculated constant-volume heat capacity $C_V$ in fixed cells.
First, we generated all possible cation configurations for the conventional unit cell of the Th$_3$P$_4$-type structure consisting of 28 atoms (4Mg$_2$GeO$_4$, 12 cation sites) with a lattice constant of 10.3 a.u., roughly corresponding to $\sim$200 GPa; the lattice vectors of this cell are $(a,0,0)$, $(0,a,0)$, and $(0,0,a)$ with $a=10.3$ a.u.. At the 12a sites, 8 Mg and 4 Ge atoms were placed. Therefore, the number of possible configurations is ${}_{12}C_8={}_{12}C_4=495$ of which only 15 are irreducible. $C_V$ calculated for this supercell is shown in Fig.~S2 and has a peak at $\sim$2100 K. Its full width at half maximum (FWHM) is considerably large and over $\sim$2000 K.

Next, we prepared a $\sqrt{2}\times\sqrt{2}\times1$ supercell consisting of 56 atoms (8Mg$_2$GeO$_4$, 24 cation sites); lattice vectors are $(a,a,0)$, $(-a,a,0)$, and $(0,0,a)$. For this supercell, the number of possible configurations is ${}_{24}C_{16}=735471$ and the number of symmetrically distinct configurations is 23253. $C_V$ calculated from 23253 configurations with distinct multiplicities has a peak at 2100 K (Fig.~S2). Its FWHM is $\approx$ 1000 K, following the expected $1/\sqrt{N}$ scaling when compared to the FWMH of the 28-atom cell calculation. However, we need to perform structural optimizations and phonon calculations at several target pressures. These calculations for all 23253 configurations are impractical. Hence, an efficient sampling of configurations is necessary. We tested the convergence of results with respect to the number of configuration as follows: for the 56-atom supercell we generated all configurations involving one, two, and three cation interchanges starting from the $I\bar{4}2d$-type structure. They produced 128, 3360, and 31360 configurations among which 4, 120, and 980 are irreducible, respectively. As the number of configurations increases, the peak in $C_V$ narrows down (see Fig.~S2). The peak temperature in $C_V{(T)}$ obtained using configuration ensembles generated by one-, two-, and three-interchanges are 2500, 2300, 2200 K, respectively, converging to that from all possible configurations, 2100 K.

Finally we prepared a supercell consisting of 224 atoms (32Mg$_2$GeO$_4$, 96 cation sites); lattice vectors are $(2a,0,0)$, $(0,2a,0)$, and $(0,0,2a)$. For this supercell, there are ${}_{96}C_{64}\sim2.97\times10^{25}$ possible cation configurations. It is very challenging to calculate $C_V$ using this number of configurations. For this supercell, there are 2048 one-interchange configurations among which 16 are symmetrically distinct. The peak in $C_V$ in this case is at 2300 K (Fig.~S2), being close to those from the configurations generated by one- and two-interchanges in the $\sqrt{2}\times\sqrt{2}\times1$ supercell. 

After comparing all $C_V$s calculated above, we chose to use the 125 configurations generated by up to two cation interchanges in the $\sqrt{2}\times\sqrt{2}\times1$ supercell to estimate $T_c$ in this ODT. We expect the resulting $T_c$ to be $\sim$200--300 K above the ideal {\it ab initio} $T_c$ prediction.

\section*{Data of cation configurations}

The archive file {\tt tau.tar} consists of {\tt tauConf$n$} ($n=001...125$), {\tt tauOxygen}, and {\tt deg.f}.
The 125 cation configurations of 8 Mg$_2$GeO$_4$ are stored in the files {\tt tauConf$n$} and {\tt tauOxygen}, archived in {\tt tau.tar}. The {\tt tauConf$n$} file contains atomic data of the $n$th cation configuration in the following format:
\begin{eqnarray*}
A_1 & \quad x_1 \quad y_1 \quad z_1 \\
A_2 & \quad x_2 \quad y_2 \quad z_2 \\
... & \\
A_{24} & \quad x_{24} \quad y_{24} \quad z_{24}, \\
\end{eqnarray*}
where $A_i$ represents atomic species, Mg or Ge, and $(x_i, y_i, z_i)$ are atomic positions in the crystal coordinates. Cartesitan coordinates are given by $x_i {\bf a}_1 + y_i{\bf a}_2 +z_i{\bf a}_3$; the lattice vectors are ${\bf a}_1=(a,a,0)$, ${\bf a}_2=(-a,a,0)$, and ${\bf a}_3=(0,0,a)$, where $a$ is the lattice constant of the conventional unit cell. The {\tt tauOxygen} file contains atomic positions of oxygen atoms in the same format. The degeneracy of the $n$th cation configuration is given in {\tt deg.f}.

\newpage
\section*{Supplementary figures}

\begin{figure}[h]
\hbox to \hsize{\hfill
\includegraphics[width=100mm]{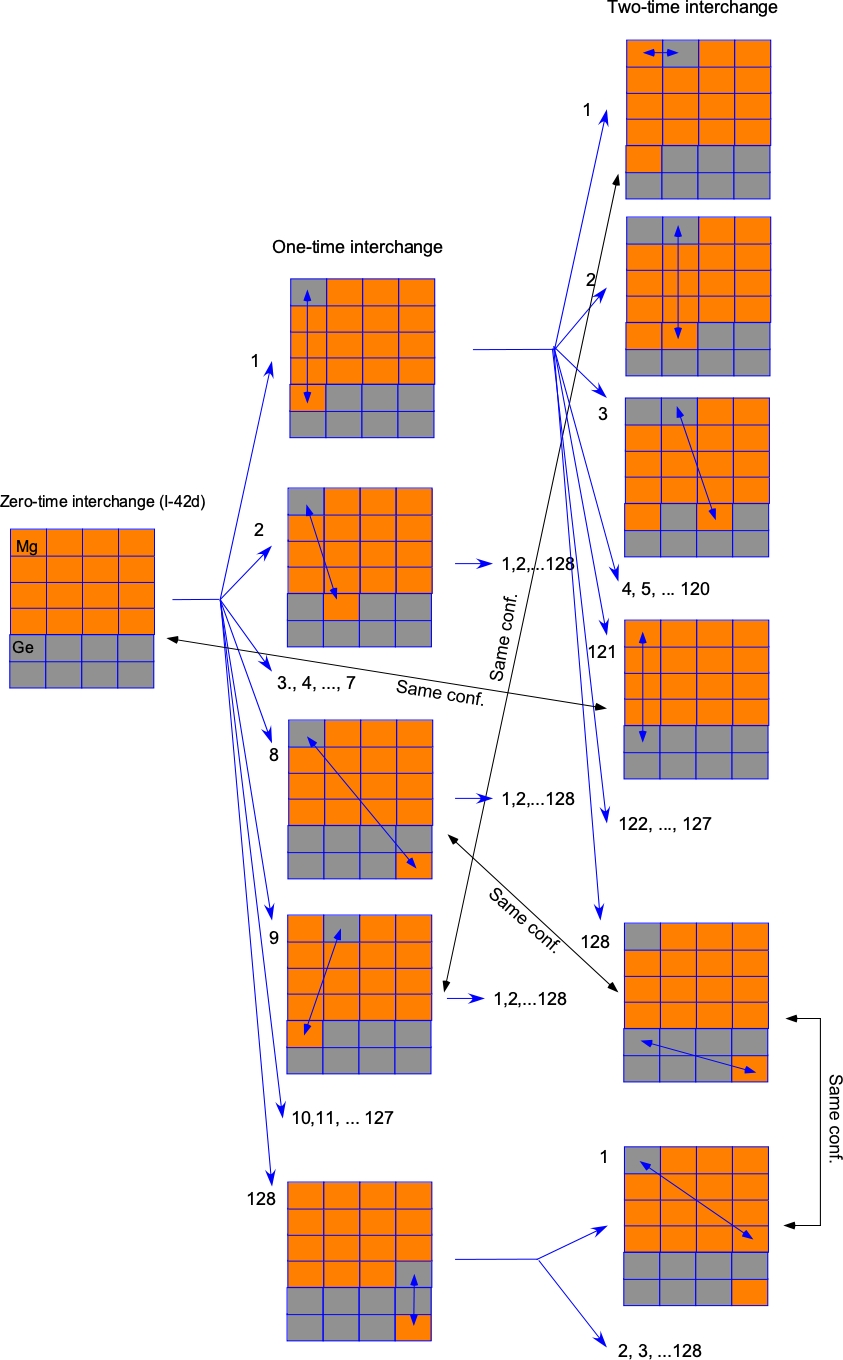}
\hfill}
\end{figure}

FIG.~S1: Diagram showing the procedure to generate one- and two-interchange cation configurations starting from the $I\bar{4}2d$-type configuration (zero-interchange).

\newpage
\begin{figure}[h]
\hbox to \hsize{\hfill
\includegraphics[width=140mm]{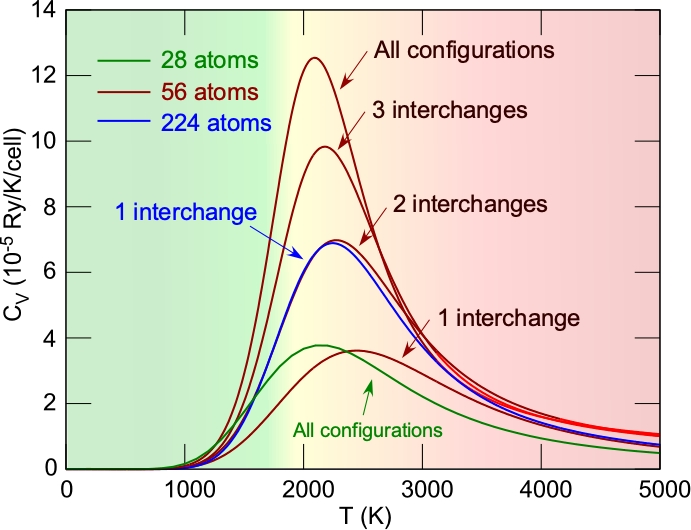}
\hfill}
\end{figure}

FIG.~S2: Constant-volume heat capacity, $C_V$, calculated in several configuration ensembles generated using different supercell sizes: all configurations in the $1\times1\times1$ conventional unit cell (green line); all configurations and configurations generated by one-, two-, and three-interchanges of cations in the 56-atom supercell (red lines); configurations from one-interchange of cations in 228-atom supercell (blue line).

\newpage
\begin{figure}[h]
\hbox to \hsize{\hfill
\includegraphics[width=140mm]{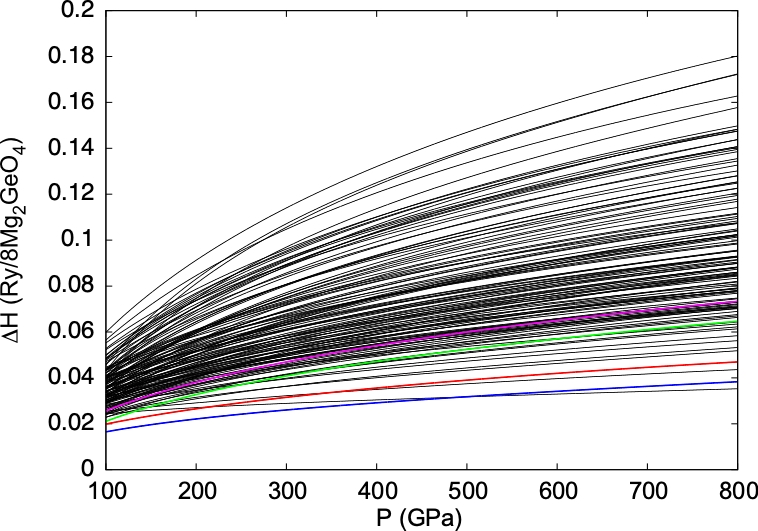}
\hfill
}
\end{figure}

FIG.~S3: Calculated enthalpy differences of all 125 configurations with respect to the $I\bar{4}2d$-type Mg$_2$GeO$_4$ in the $\sqrt{2}\times\sqrt{2}\times1$ supercell. Colored and black lines represent configurations obtained with one- and two-interchange of cations, respectively. Among the 125 configurations, the $I\bar{4}2d$-type phase has the lowest enthalpy at all pressures investigated here. The 4 configurations generated by one-interchange of cations tend to have lower enthalpies (colored lines) than the 120 configurations by two-time interchanges (black lines), with some exceptions.